\documentstyle[12pt,aaspp4]{article}
\def \yskip{\penalty-50\vskip3pt plus 3pt minus 2pt}
\def \pp{\par \yskip \noindent \hangindent .4in \hangafter 1}
\def \abc#1#2#3#4 {\pp#1, {\sl#2}, {\bf#3}, #4}
\def \blank {\lower 5pt\hbox to 0.75in{\hrulefill}}

\newfont{\rten}{cmr10}

\begin{document}

\normalsize
 
\title{The Near-Infrared Galaxy Counts Anomaly: Local Underdensity or
Strong Evolution?}
\vspace*{0.3cm}

\author{L. A. Phillips and Edwin L. Turner}
\affil{Princeton University Observatory, Princeton, NJ 08544}
\affil{email: phillips, elt@astro.princeton.edu}

\begin{abstract}

We analyze bright-end ($K = 10 -17$) galaxy 
counts from a number of near-infrared galaxy surveys.
All studies available as of mid-1997, considered individually or collectively, show that the 
observed near-infrared galaxy number counts at 
low magnitudes are inconsistent with a simple no-evolution model.
We examine evolutionary effects and a local underdensity model 
as possible causes of this effect. We find that the data are fit by either a factor of 
$1.7$ - $2.4$ deficiency of galaxies out to a redshift of 
$z = 0.10$ - $0.23$, depending on the k corrections and 
evolution (e-) corrections used and the adopted values of the Schechter 
luminosity function parameters, or by unexpectedly strong low redshift
evolution in the K-band, leading to (e+k)-corrections at $z= 0.5$ that 
are as much as 60\% larger than accepted values.  The former possibility
would imply that the local expansion rate on scales of several hundred
Mpc exceeds the global value of $H_0$ by up to 30\% and that the amplitude
of very large scale density fluctuations is far larger than expected in
any current cosmogonic scenario.  The latter possibility would mean that
even the apparently most secure aspects of our understanding of galaxy
evolution and spectral energy distributions are seriously flawed.

\end{abstract}

\begin{keywords}
{cosmology: miscellaneous --- galaxies: evolution --- infrared: galaxies --- 
large-scale structure of the universe}
\end{keywords}

\clearpage

\section{Introduction}

The past ten years have seen a proliferation of near-infrared galaxy
surveys providing us with increasingly accurate K-band galaxy counts.
The surveys have covered both the bright end (Gardner et al. 1996, 
Gard96 hereafter; Huang et al. 1997, Huang97 hereafter; Glazebrook et al. 1994; 
Gardner, Cowie, \& Wainscoat 1993; Mobasher, Ellis \& Sharples 1986) 
and faint magnitudes (Moustakas et al. 1997,
Djorgovski et al. 1995; McLoed et al. 1995; Soifer et al. 1994;
Jenkins \& Reid 1991.) 

Galaxy counts are an essential tool in the study of galaxy
evolution and cosmological geometry. The advantage of
near-infrared galaxy counts over optical counts is that
the former are much less sensitive to stellar population evolution
and the K-band k-corrections are smoother and better understood
than in the bluer region of the spectrum. 
In particular, they are much less affected by bursts of star formation
and internal extinction by dust, K-band galaxy
counts are therefore the ideal tool for probing large scale density variations. 

 Huang97 report that the slope of
their bright-end galaxy counts is steeper than that predicted by a 
no-evolution model. The authors dismiss known 
observational error and evolution as likely causes 
for this effect and describe a heuristic model with a local deficiency of 
galaxies by a factor of 2 on scales sizes of around 300 $h^{-1} Mpc$
as a possible cause. But Gard96 find that this model 
does not fit their galaxy counts.

In this paper, we provide a more in depth study of bright-end galaxy 
counts in order to determine if other published
 galaxy counts point to a similar local deficiency of galaxies and
 to obtain a more precise picture of this underdensity using all 
K band galaxy counts available as of mid-1997.
The possible existence of a very large scale local underdensity merits
close scrutiny since it could have profound implications for the 
determination of $H_{0}$ (Turner, Cen, \& Ostriker, 1992.)
 We also attempt to quantify the luminosity evolution that would be needed to account 
for the count slope discrepancy, the most plausible alternative explanation.
 The data used, models, and best fit methods are described 
in sections 2, 3, and 4 respectively. Results are presented in section 5
and discussed in section 6.

\section{Data}

Individual fits were done to the data from Huang97 and 
Gard96. In order to insure good coverage of the magnitude
range under consideration, the rest of the data, listed in Gardner et al.
(1993), were combined. Fits with those and with all the data combined
were also calculated. The magnitude ranges as well as the total areas 
covered by the different surveys are listed in Table 1.
 
A compilation of bright-end number counts is shown in Figure 1a. 
Density evolution would appear as an excess in the number counts 
over the no-evolution
model. As Figure 1b shows, the slope of the combined low magnitude galaxy counts
 is indeed steeper than that of the no-evolution model 
($d\log(N)/dm = 0.67 \pm 0.01$,
up to $K = 15$ compared with the Euclidean value of $0.6$.)

\section{Models}
\subsection{Galaxy Counts}
The expression for the differential number count, the number of 
galaxies per degree$^2$ per magnitude as a function 
of apparent magnitude is (Yoshii \& Takahara 1988, hereafter YT88)
\begin{eqnarray}
N(m)  & = &  \int_{0}^{z_{f}} n(m,z)dz  \\ 
n(m,z) &  = & \frac{\omega}{4\pi} \frac{dV}{dz} 
\sum_{i=1}^{5}\psi^{i}(M)\ ; 
\end{eqnarray}
where $\omega$ is the area is steradians. For the near-infrared, the
sum over the different galaxy types can be approximated by the
Schechter luminosity function (Schechter 1976)
\begin{eqnarray}
\sum_{i=1}^{5}\psi^{i}(M)  & \simeq  & \phi (L)dL \nonumber \\
& =  & \phi^{*} (L/L^{*})^\alpha \exp(-L/L^*) d(L/L^{*})\\  
& =  & \ln A \phi^{*}\exp\{- \ln A (\alpha+1)
(M-M^{*})-\exp[- \ln A (M-M^{*})\} dM\ ,
\end{eqnarray}
from YT88, with $\ln A = 0.4 \ln 10$ and
\begin{equation}
M = m - K(z) - E(z) -5\log_{10} (d_{L}/10pc)\ ,
\end{equation}
where $M$, $m$, $K(z)$ and $E(z)$ are the absolute magnitude,
the apparent magnitude, the k-correction and the evolution
correction, respectively.
The co-moving volume is (YT88)
\begin{equation}
\frac{dV}{dz} = 
\frac{4\pi c\,d_{L}^{2}}{H_{0}(1+z)^{3}\sqrt{1+2q_{0}z}}\ ,
\end{equation}
and, in the Friedman model with $\Lambda = 0$, the luminosity 
distance is (YT88)
\begin{equation}
d_{L} = 
\frac{c}{H_{0}q_{0}^{2}} \{q_{0}z +(q_{0}-1)
(\sqrt{1+2q_{0}z} -1)\}\ , 
\end{equation}
where $H_{0}$, $q_{0}$, and c are the Hubble constant, the
deceleration parameter and the light velocity, respectively.

With $E(z)$ set to zero, this is the no
evolution model. The $H_{0}$ value cancels out in the above equations
and $q_{0} = 0.5$, except when otherwise indicated. Varying the deceleration parameter 
does not affect our fits which use counts only up to $K=17$ where 
the model is negligibly affected by cosmology. 

Normally $z_{f} \simeq 5$ (YT88) but since we are only using data
with $K\leq 17$, we integrate only to $z_{f} = 1.5$ where we still
know the k-correction well (Huang97; 
Glazebrook et al. 1995.) The redshift distribution of galaxies with 
$K = 17$ tapers off well before this redshift both observationally
(Gardner et al. 1998) and in our model.
  
Our results are based primarily on Schechter luminosity function parameters calculated by 
Gardner et al. (1997) ($M^{*}=-23.12 +5\log h$, $\phi^{*} =1.66\times 10^{-2} h^3 Mpc^{-3}$ 
and $\alpha =-0.91$, where $h=H_{0}/100$.)  
As alternate possibilities, we consider  Mobasher et al. (1993) 
($M^{*}=-23.59+5\log h$, $\phi^{*}=0.14\times 10^{-2}(H_{0}/50)^3 Mpc^{-3}$, 
and $\alpha =-1.0$) and Glazebrook et al. (1995) ($M^{*}=-22.75 + 5\log h$, 
$\phi^{*} = 0.026h^3 Mpc^{-3}$ and $\alpha =-1.0$)
Schechter luminosity function parameters, to test for the sensitivity of our results
to these parameters.

K-corrections derived from both the 'UV-hot' elliptical model
(Rocca-Volmerange \& Guiderdoni 1988) 
and evolutionary synthesis models (Bruzual \& Charlot 1993)
are used in our model in order to bracket the range of plausible K(z) terms.
We use (e+k)-corrections for $\Omega = 1.0$ and $0.2$, 
computed as a weighted average of the corrections given in 
Poggianti (1997) for elliptical, Sa and Sc galaxies. 
The contributions from different Hubble types we use for these
'Averaged' (e+k)-corrections is 30\%E /20\%Sa /50\%Sc
(van den Berg et al. 1996.) Similarly averaged 
(e+k)-corrections calculated by Pozzetti, Bruzual, \& Zamorani (1996) for 
 $\Omega \sim 0$ are also used.

In addition, we use (e+k)-corrections for the Hubble type
that has the largest such corrections as a conservative 
limiting case in our fits to obtain a lower limit on the 
size of the required local underdensity in the presence of evolution. 
Pozzetti et al. (1996) find elliptical galaxies to have the largest K-band 
(e+k)-corrections. Poggianti (1997) find Sa galaxies to have only slightly
bigger (e+k)-corrections than ellipticals. We therefore use (e+k)-corrections 
for ellipticals calculated (for $\Omega \sim 0$) by Pozzetti et al. (1996) 
using a pure luminosity evolution model, and for Sa galaxies calculated 
(for $\Omega = 1.0$ and $0.2$) by Poggianti (1997) using an evolutionary
synthesis model (Poggianti \& Barbaro 1996.) 

\subsection{Underdensity}

In order to introduce a local underdensity in our model, we
substituted
\begin{equation}
\phi^{*} \longrightarrow \phi^{*}D(z)\ , 
\end{equation}
with D(z) taking one of the following two forms: the 'step' underdensity
\begin{equation}
D(z) = \cases{1.0 - d ,& if $z \leq w$ \cr
1.0 ,& if $z > w$} \ ,
\end{equation}
where d and w are the depth and width fit parameters of the
underdensity, and the 'smooth' underdensity
\begin{equation}
D(z) = \frac{d}{[d+(1-d)\exp\{-(z/w)^2\}]} \ ,
\end{equation}
which is a generalization of the underdensity proposed by Huang97
 and again d, w and $\phi^{*}$ are fit parameters.

\subsection{Strong Evolution}
In order to quantify the evolution needed to account for the steep slope
if there is no local underdensity, we assume that luminosity evolves as a power of time 
and substitute the following function into equation (5)
\begin{equation}
E(z) = -a\log_{10} (1+z)\ ,
\end{equation}
where $a$ is a fit parameter. We also try the following function,
assuming that the evolution ceased some time ago,
\begin{equation}
E(z) = \cases{0.0, & if $z \leq b$ \cr
-a\log_{10}(1+(z-b)),& if $z > b$} \ ,
\end{equation}
where $a$ and $b$ are fit parameters.
\section{Techniques}

\subsection{Best Fit Methods}
We used two methods to determine the best fit parameters for both 
the strong evolution and the underdensity models. 
The first was $\chi ^2$ minimization. We simply minimized
the usual expression in which we used averaged logarithmic error bars,  
$\sigma \equiv [\sigma_{upper} + \sigma_{lower}]/2$. The logarithm of the
galaxy counts and corresponding error bars are used since the latter
 are more symmetric than the non-logarithmic error bars. This
expression unfortunately precludes the use of data points whose error bars overlap zero 
and information may be lost by averaging the error bars.
However, this method does have the advantage of allowing for an estimate of the
goodness-of-fit by looking at the probability that a $\chi^{2}$ 
as poor as the value obtained should occur by chance 
\begin{equation}
Q = \frac{1}{\Gamma(a)} \int_{x}^{\infty} e^{-t}t^{a -1}\,dt\ ,
\end{equation}
where $a = \frac{\nu}{2}$ with $\nu =$ number of degrees of
freedom, and $x = \frac{\chi^{2}}{2}$.
 As a rule, $Q \ge 0.1$ indicates a believable goodness-of-fit but when, 
as will be the case, errors 
are non-normal, a $Q$ as small as $0.001$ is acceptable (Press
et al. 1992) 

The second fitting method we use is the Poissonian maximum 
likelihood function which 
has the form
\begin{equation}
ML = \prod_{i=1}^{n} \frac{\lambda_{i}^{X_{i}}
\exp{(-\lambda_{i})}}{X_{i}!}\ ,
\end{equation}
where $X_{i} \equiv$ (number count) $\times$ (area)
and $\lambda_{i} \equiv$ (model number count) $\times$ (area)
and the area is that of the survey from which this particular
galaxy number count data is taken. 
We maximize the natural logarithm of equation (14)
\begin{equation}
\log{ML} = \sum_{i=1}^{n}X_{i}\log{\lambda_{i}} -
\sum_{i=1}^{n}\lambda_{i} - \sum_{i=1}^{n}\log X_{i}!\ ,
\end{equation}

The assumption that the errors are well approximated by Poisson 
statistics holds for the Huang97 data, as the error 
analysis in that article suggests, as well as for data from the 
other surveys used in our fits, except at the very faintest 
magnitudes of the HMWS and HMDS data where uncertainty in 
star/galaxy separation becomes important (Gardner 1992.) 

\subsection{Error Determination}

We use Monte-Carlo simulations to determine the errorbars for
our underdensity fits. We generate random Gaussian deviates
with zero mean and unit variance and set
\begin{equation}
N(m) = \cases{N_{0}(m) + \sigma_{lower} \times RGD, & if $RGD \leq 0.0$ \cr
N_{0}(m) + \sigma_{upper} \times RGD, & if $RGD > 0.0$} \ ,
\end{equation}
where $N_{0}(m)$ is the actual galaxy count from the data
and $RGD$ is a random Gaussian deviate.  The simulated data 
sets are then run through the best fit algorithms and 
the results are used to calculate a variance 
from the best fit results to the actual data. For the 'step' underdensity 
model, the errorbars can be expressed as 
\begin{equation}
\sigma = \left[ \frac{\sum_{i=1}^{N}(p_{0} - p_{i})^{2}}{N}
\right]^{1/2}\ ,
\end{equation}
where $N$ is the number of simulated data sets, $p_{0}$ and
$p_{i}$ are parameters resulting from the fit to the actual data
 and simulated data, respectively.

We illustrate the distribution of underdensity
values for the 'smooth' underdensity fits by dividing the 2-dimensional 
$\phi^{*}D(z)$ - $z$ space into cells and calculating the probability of 
the simulated data passing through that particular cell. The 
probability distribution is then plotted as a grey scale plot.

\section{Results}

The results of the fits using Schechter parameters $M^{*}$ and
$\alpha$  from Gardner et al. 
(1997) and the 'Step' underdensity described by equation (10)
are listed in Table 2. The densities at the center of, and outside the
 underdensity are $\phi^{*}_{center}$ and 
$\phi^{*}_{out}$, respectively. The last column shows the percentage difference 
between these two densities. The 2$\sigma$ error bars (95\% confidence level)
given in this and subsequent tables were obtained using Monte-Carlo simulations.
Only results from the maximum likelihood fits are presented here. Results 
from best $\chi^{2}$ fits differ by at most 0.5$\sigma$ for the individual fits 
(including the fits to 'Other data') and by 0.9$\sigma$ at most for 
the fits to all the data combined, except for the last eight fits in the table
(with the Poggianti (1997) $\Omega =0.2$ and the Pozzetti et al. (1996)
 $\Omega \sim 0$ (e+k)-corrections,) where these numbers 
become 0.9$\sigma$ and 1.3$\sigma$, respectively. The $Q$ values in 
column 2, are therefore presented only as an indication of the goodness of fit.

For a given k-correction, the results for all the individual
data sets (Gard96, Huang97, and 'Other data',
for the rest of this discussion) are consistent with each other.
 Results from most fits to individual data sets differ from the
fit to all the data combined by less than 2$\sigma$ and all have $Q > 0.1$.
The two fits to the Gard96 data using Poggianti (1997) corrections, 
however, differ by as much as 2.4$\sigma$ for the $\Omega = 1$ fit and by as much as 5$\sigma$ 
for the $\Omega = 0.2$ fit. The fits to all the data combined are at least
6$\sigma$ results and have $Q > 0.01$. The results in Table 2
point to a factor of $\sim 2$ deficiency in galaxies out to redshifts
of $z = 0.12 - 0.18$.

Only the results from fits to all the data combined are
presented here for Schechter parameters $M^{*}$ and
$\alpha$ calculated by Glazebrook et
al (1995), in the first half of Table 3, and for those calculated by
Mobasher et al (1993), in the second half of Table 3. All the
results have $Q > 0.01$ and point to a factor of 1.7 - 2.5
deficiency of galaxies out to redshifts of $z = 0.09 - 0.23$
with at least a 99.99\% confidence level.

The fits using limiting cases of published (e+k)-corrections, yielded underdensities that were
significant to at least the $5\sigma$ level. The fits using (e+k)-corrections for Sa galaxies only 
from Poggianti (1997) yielded a factor of 1.6 - 1.9 deficiency of galaxies 
out to a redshift of $z = 0.08 - 0.15$.  The underdensities obtained in fits using 
(e+k)-corrections from Pozzetti et al. (1996) for ellipticals only and $\Omega \sim 0$
had a factor of 1.4 - 2.0 deficiency in galaxies out to $z = 0.11 - 0.24$.  

The 'Smooth' Underdensity results for maximum likelihood 
fits to all the data combined,
again for $M^{*}$ and $\alpha$ calculated by Gardner et al. (1997)
are shown as grey scale plots in Figure 3. 
The darker regions indicate a greater probability 
of the fit yielding a density profile that passes through that
particular point. The probability distribution was calculated from 
Monte-Carlo simulations as described in section 4.2. 
The solid lines represent the actual fit to the 
data. Plots for Bruzual \& Charlot (1993), in (a), and 
Rocca-Volmerange \& Guiderdoni (1988), in (b), k-corrections are shown.
Plots for the 'Average' Poggianti (1997) (e+k)- corrections for $\Omega = 1.0$
and $0.2$ are shown in Figure 3c and 3d, respectively. These fits to all the
data combined all have $Q > 0.01.$

The mean redshifts of the redshift distributions for $K = 14.0$ and $K=17.0$
are $z = 0.095$ and $z = 0.32$ in the absence of k- or evolution corrections and shift to 
$z = 0.11$ to $z = 0.47$ when (e+k)-corrections (Poggianti 1997) are included. 
The redshift range involved in the determination of the underdensity 
parameters is therefore well sampled by the data, regardless of the 
e- and k-correction used. The width and depth of the underdensity 
are therefore not uniquely determined by the faintest
magnitude bins and it is unlikely that the observed 
underdensity is in fact an artifact of Malmquist bias. 
We also tried allowing only $\phi^{*}$ to vary and not
introducing an underdensity. This lead to best fit results with 
 $Q \ll 1.0 \times 10^{-4}$ and maximum likelihood values
at least a factor of $10^{22}$ times smaller than those 
obtained for the underdensity fits.
 
The evolution model described by equation (11) 
did not yield satisfactory fits. We tested the model using
Schechter parameters determined by Gardner et al. (1997) 
using both Bruzual \& Charlot (1993) and Rocca-Volmerange \& Guiderdoni (1988)
k-corrections and obtained values of $Q \le  0.001$ for most fits,
 indicating a poor goodness-of-fit. Maximum likelihood values 
were a factor of $10^{4.8}-10^{84}$ smaller than those 
obtained in the underdensity fits with the same Schechter
parameters and k-corrections.

Substantially better results were obtained for the strong evolution model
described by equation (12). The best fits, obtained using 
the Bruzual \& Charlot (1993) k-corrections are shown in Figure 4. 
However, both Q and maximum likelihood values are again systematically
lower (by a factor of $10^{0.8}-10^{21.4}$ for the maximum
likelihood) than those obtained in the underdensity fits. 
The poorer fits obtained with the evolution models may reflect
the fact that the effects of evolution are small at the bright 
magnitudes, and hence overall low redshifts, that are of interest 
in modeling the steep slope.

\section{Discussion}

Turner et al. (1992) found an approximate expression of 
the correction from the local to global $H_{0}$, $(\Delta
H_{0}/H_{0}) = -0.6\times \Delta n_{gal} \Omega^{0.4}$, where
$\Delta n_{gal}$ is the over or under density within the local
volume. This would lead, in our case to a local evaluation of 
$H_{0}$ that is as much as $30\%$ (for $\Omega \leq 1$) higher 
than the global value. However the authors warn that the presence of coherent
structure with sizes $> 10000$ km/s might lead to a more
extreme effect.

Recently, Kim et al. (1997) have shown that $(\Delta H_{0}/H_{0}) <
0.10$ if $\Omega_{M} \leq 1$, using seven supernovae with 
$0.35 < z < 0.65$. However the authors mention that errors in
absolute magnitude calibrations could affect this ratio quite 
strongly, pointing to the 0.09 mag difference between the
absolute magnitude calibrations used in their paper and by 
Riess, Press, \& Kirshner (1996) which could lead to a ratio of
 $(\Delta H_{0}/H_{0}) < 0.21$. The possibility that the local 
and the global values of $H_{0}$ differ by of order 20\% 
cannot yet be ruled out directly.

 Zehavi et al. (1998) have analyzed the peculiar velocities of 44 
Type Ia supernovae and found a deviation from the Hubble law consistent with
a void of $\sim 20\%$ underdensity surrounded by a dense wall at $70h^{-1} Mpc.$
This result is consistent with those obtained from peculiar 
motions of rich clusters (Lauer \& Postman 1992; Lauer et al. 1997) but
cannot be used to explain the steep slope of near-infrared galaxy count. 
With this small local void introduced in our model, our underdensity fits
yield voids that are at most a factor of 1.15 smaller in extent, but
as much as 1.25 times more deficient in galaxies, than those listed 
in Tables 2 and 3.

However, Wang, Spergel, \& Turner (1998) used current knowledge of CMB anisotropies 
to show that a variation of a few percent between available measurements of $H_{0}$
and its true global value should be expected and that a variation as large as $10\%$
would be possible for surveys with diameter $200h^{-1} Mpc$. For larger surveys scales,
not much variation is expected. For example, for a survey with a
$500h^{-1} Mpc$ radius, variations of at most $2\%$ in expansion parameter
and $13\%$ in density are expected at the $95\%$ confidence level.
This would seem to mitigate against the large scale
underdensity found by our fits if current cosmogonic theories are at least
roughly valid.

The goodness of fit ($Q > 0.01$) and 
maximum likelihoods values obtained
with the second evolution model presented in 
section 3.3 seem to indicate that this is a viable alternative to the
underdensity models. However, as we can see in Figure 4,
our fit results show more evolution than can be accounted for 
by present evolution calculations. In fact, the (e+k)-correction from the 
fits is as much as 56\% stronger than that of the 'Averaged' 
(e+k)-corrections for $\Omega = 0.2$  used 
in our underdensity fits (for example, at $z = 0.5$, $\Delta M_{(e+k)} = -1.12$
compared with $-0.74$) and 34\% stronger than even the (e+k)-corrections
for Sa galaxies alone (again for $\Omega = 0.2$) which we took to represent an 
upper limit on the evolution (and for which $\Delta M_{(e+k)} = -0.91$ at $z = 0.5$.) 

A more careful consideration 
of evolution as a source of the effect noticed by Huang97 
and confirmed by our fits for all the data is necessary since 
our understanding of galaxy evolution is still incomplete.
The recent publication of an extensive data base for Galaxy Evolution 
Modeling (Leitherer et al. 1996) might lead to larger 
(e+k)-corrections in the near-infrared.
 A model including both evolution and an underdensity might prove to be a more
acceptable solution than either one alone, but cannot yet be
studied with the data presently available. 
Redshift surveys should ultimately answer the question of  
whether or not a large region of the local universe is underdense.

However, for the present, the K-band galaxy counts pose a significant
puzzle:  unless several independent determinations are giving similar
but incorrect results, we must confront the possibility of either a cosmic
density fluctuation of entirely unanticipated scale and amplitude or a
serious deficiency in the best understood features of galaxy evolution and
spectral energy distributions (or, of course, some combination of the two).

Acknowledgments

We are grateful to M. Fukugita and P. Garnavich for their help
and useful comments. This research was supported by NSF grant
AST94-19400.

\clearpage

\begin{center}
Figure Captions
\end{center}

Figure 1. (a) Compilation of bright end near-infrared galaxy counts taken from 
Gardner et al. (1996), Huang et al. (1997) and the Gardner et al. (1993) compilation (number counts 
for Glazebrook et al. (1994) and Mobasher et al. (1986)  are listed in this paper 
as well.) The solid line represents the actual slope of the combined galaxy 
counts up to $K = 15$ and the doted line shows the Euclidean slope. (b) Same as (a),
with the Euclidean slope $d\log(n)/dm = 0.6$ scaled out and only points with
fractional error $\leq 0.5$ and $K \leq 15$ plotted.

Figure 2. 'Smooth' underdensity fit results for all the data using
 $M^{*}=-23.12 +5\log h$ and $\alpha =-0.91$ and (a) Bruzual \&
Charlot (1993) k-corrections, (b) Rocca-Volmerange \& Guiderdoni (1988)
k-corrections and averaged (e+k)-corrections from
Poggianti (1997) for (c) $\Omega = 1.0$ and (d) $\Omega = 0.2$. 
Darker regions indicate a greater probability 
of the fit yielding a density profile that passes through that
particular point. The solid lines represent the best fit to the 
data.
 
Figure 3. Evolution best fit results with $M^* =-23.12 +5\log h$, 
$\alpha =-0.91$, Bruzual \& Charlot (1993) and Rocca-Volmerange \&
Guiderdoni (1988) k-corrections (shown in solid line
and labeled $K(z)_{BC}$ and $K(z)_{RG}$, respectively.) The
Average and Sa (e+k)-corrections from Poggianti (1997) are also shown.
 The corresponding $\Omega$ is indicated on the right.

\clearpage

\makeatletter
\def\jnl@aj{AJ}
\ifx\revtex@jnl\jnl@aj\let\tablebreak=\nl\fi
\makeatother

\small 
\begin{deluxetable}{llr}
\tablewidth{25pc}
\tablecaption{Sources of Near Infrared Galaxy Counts}
\tablehead{
\colhead{Survey} 
& \colhead{Range \tablenotemark{a}} 
& \colhead{Area \tablenotemark{b}} 
}
\startdata
Gardner et al. 1996 & 10.25 to 15.75 &   9.84 $degree^{2}$\nl
Huang et al. 1997	& 11.25 to 14.50 &   9.81 $degree^{2}$\nl
		& 14.75 to 16.00 &   8.23 $degree^{2}$ \nl
Other data \tablenotemark{c}   &        &             \nl
HWS             & 10.50 to 14.50 &   1.58 $degree^{2}$  \nl
HMWS            & 12.75 to 16.75 & 582.01 $arcmin^{2}$ \nl
HMDS            & 13.75 to 18.75 & 167.68 $arcmin^{2}$  \nl
Glazebrook et al. 1994 & 13.50 to 16.50 & 551.90 $arcmin^{2}$\nl
Mobasher et al. 1986 & 10.25 to 12.25 &  41.56 $degree^{2}$\nl
\tablenotetext{a}{of data in $K$ magnitude}
\tablenotetext{b}{Total area covered for $K \le 17$}
\tablenotetext{c}{Listed in Gardner et al. 1993}
\enddata
\end{deluxetable}

\clearpage

\small
\begin{deluxetable}{lrllll}
\tablecaption{Results of 'Step' Underdensity Fit 
($M^* =-23.12 +5\log h$, $\alpha =-0.91$)}
\tablehead{ 
\colhead{Survey}
& \colhead{Q/$\ln ML$}
& \colhead{$\phi^{*}_{center} (\times 10^{-2})$}
& \colhead{$\phi^{*}_{out} (\times 10^{-2})$}
& \colhead{Width (z)}
& \colhead{$\Delta n_{gal}$ (in \%)}
} 
\startdata
\multicolumn{6}{c}{Bruzual \& Charlot K-corrections}\nl
Gardner et al. & 0.614/-39.20 & 1.82 $\pm$ 0.58 & 3.06 $\pm$ 0.76 
& 0.220 $\pm$ 0.083 & 40.5 $\pm$ 11.8\nl
Huang et al. & 0.32/-76.58 & 1.40 $\pm$ 0.41 & 2.90 $\pm$ 0.33 
& 0.146 $\pm$ 0.044 & 51.6 $\pm$ 13.2\nl
Other data \tablenotemark{a} & 0.288/-109.06 & 1.20 $\pm$ 0.54 & 2.71 $\pm$ 0.54 
& 0.188 $\pm$ 0.058 & 55.6 $\pm$ 18.0\nl
All data & 0.0704/-265.38 & 1.48 $\pm$ 0.23 & 2.87 $\pm$ 0.20 
& 0.156 $\pm$ 0.032 & 48.5 $\pm$ 7.4\nl
\multicolumn{6}{c}{Rocca-Volmerange \& Guiderdoni K-corrections}\nl
Gardner et al. & 0.462/-40.61 & 1.46 $\pm$ 0.50 & 2.70 $\pm$ 0.68 
& 0.259 $\pm$ 0.102 & 45.9 $\pm$ 12.8 \nl
Huang et al. & 0.354/-75.63 & 1.13 $\pm$ 0.35 & 2.43 $\pm$ 0.31 
& 0.170 $\pm$ 0.058 & 53.6 $\pm$ 13.3\nl
Other data & 0.274/-109.40 & 1.02 $\pm$ 0.47 & 2.32 $\pm$ 0.44 
& 0.221 $\pm$ 0.038 & 56.2 $\pm$ 18.4\nl
All data & 0.0803/-262.85 & 1.21 $\pm$ 0.18 & 2.42 $\pm$ 0.20
& 0.183 $\pm$ 0.038 & 49.9  $\pm$ 6.6\nl
\multicolumn{6}{c}{Poggianti E+K-corrections ($\Omega = 1.0$)}\nl
Gardner et al. & 0.658/-38.87 & 1.82 $\pm$ 0.39 & 2.96 $\pm$ 0.42 
& 0.190 $\pm$ 0.050  & 38.6 $\pm$  9.72\nl
Huang et al. & 0.141/-89.60 & 1.30 $\pm$ 0.55 & 2.87 $\pm$ 0.38 
& 0.124 $\pm$ 0.062 & 54.7 $\pm$ 18.2\nl
Other data & 0.250/-109.82 & 1.20 $\pm$ 0.53 & 2.51 $\pm$ 0.45 
& 0.162 $\pm$ 0.082 & 52.1 $\pm$ 19.1 \nl
All data & 0.0267/-280.74 & 1.42 $\pm$ 0.27 & 2.84 $\pm$ 0.22 
& 0.134 $\pm$ 0.041 & 50.2 $\pm$  8.7 \nl
\multicolumn{6}{c}{Poggianti E+K-corrections ($\Omega = 0.2$)}\nl
Gardner et al. & 0.649/-38.93 & 1.82 $\pm$ 0.33  & 2.84 $\pm$ 0.28 
& 0.180 $\pm$ 0.018 & 35.7 $\pm$ 9.8 \nl
Huang et al. & 0.108/-92.45 & 1.27 $\pm$ 0.47 & 2.76 $\pm$ 0.30  
& 0.116 $\pm$ 0.033 & 54.0 $\pm$ 16.1\nl
Other data & 0.237/-110.08 & 1.20 $\pm$ 0.55 & 2.35 $\pm$ 0.47 
& 0.151 $\pm$ 0.112 & 48.9 $\pm$ 21.0 \nl
All data & 0.0186/-285.56 & 1.40 $\pm$ 0.21 & 2.73 $\pm$ 0.17 
& 0.124 $\pm$ 0.022 & 48.5 $\pm$ 7.14 \nl
\multicolumn{6}{c}{Pozzetti et al. E+K-corrections ($\Omega \sim 0$)}\nl
Gardner et al. & 0.628 /-39.13 & 1.86 $\pm$ 0.42 & 3.52 $\pm$ 1.12 
& 0.223 $\pm$ 0.108 & 47.3 $\pm$ 16.5 \nl
Huang et al. & 0.353/-77.73 & 1.44 $\pm$ 0.23 & 3.25 $\pm$ 0.34  
& 0.157 $\pm$ 0.016  & 55.6 $\pm$ 13.4 \nl
Other data & 0.292/-108.9 & 1.23 $\pm$ 0.34 & 3.11 $\pm$ 0.69
& 0.194 $\pm$ 0.037 & 60.4 $\pm$ 19.8 \nl
All data & 0.0744/-266.1 & 1.52 $\pm$ 0.10 & 3.08 $\pm$ 0.18
& 0.154 $\pm$ 0.017 & 50.6 $\pm$ 6.1\nl
\tablenotetext{a}{Listed in Table 1}
\enddata
\end{deluxetable}

\clearpage

\small
\begin{deluxetable}{lrllll}
\tablecaption{Results of 'Step' Underdensity Fit 
for all data combined using other $M^{*}$ and $\alpha$}
\tablehead{ 
\colhead{K(z) \& E(z) \tablenotemark{a}}
& \colhead{Q/$\ln ML$}
& \colhead{$\phi^{*}_{center} (\times 10^{-2})$}
& \colhead{$\phi^{*}_{out} (\times 10^{-2})$}
& \colhead{Width (z)}
& \colhead{$\Delta n_{gal}$ (in \%)}
} 

\startdata
\multicolumn{6}{c}{$M^* =-22.75 +5\log h$, $\alpha =-1.0$ \tablenotemark{b}}\nl
(1) & 0.0444/-274.08 & 2.40 $\pm$ 0.39 & 4.25 $\pm$ 0.29 
& 0.104 $\pm$ 0.030 & 43.6 $\pm$ 8.4\nl
(2) & 0.0509/-271.13 & 2.07 $\pm$ 0.28 & 3.54 $\pm$ 0.18 
& 0.132 $\pm$ 0.009 & 41.6 $\pm$ 7.2 \nl
(3) & 0.0186/-285.63 & 2.36 $\pm$ 0.44 & 4.41 $\pm$  0.34
& 0.099 $\pm$ 0.033 & 46.6 $\pm$ 9.1 \nl
(4) & 0.0131/-290.23 & 2.33 $\pm$ 0.37 & 4.29 $\pm$ 0.24
& 0.093 $\pm$ 0.016 & 45.8 $\pm$ 8.1 \nl
(5) & 0.0542/-268.5 & 2.49 $\pm$ 0.28 & 4.64 $\pm$ 0.72
& 0.115 $\pm$ 0.067 & 46.3 $\pm$ 8.7 \nl
\multicolumn{6}{c}{$M^* =-23.59 +5\log h$, $\alpha =-1.0$ \tablenotemark{c}}\nl
(1) & 0.0701/-266.36 & 0.768 $\pm$ 0.160 & 1.77 $\pm$ 0.25 
& 0.204 $\pm$ 0.069 & 56.7 $\pm$ 6.6 \nl
(2) & 0.0938/-261.73 & 0.624 $\pm$ 0.115 & 1.52 $\pm$ 0.17
& 0.233 $\pm$ 0.054 & 58.9 $\pm$ 6.0 \nl
(3) & 0.0300/-279.10 & 0.750 $\pm$ 0.11 & 1.64 $\pm$ 0.11
& 0.171 $\pm$ 0.027  & 54.4 $\pm$ 6.1 \nl
(4) & 0.0187/-285.28 & 0.741 $\pm$ 0.11 & 1.55 $\pm$ 0.10 
& 0.156 $\pm$ 0.025 & 52.3 $\pm$ 6.7 \nl
(5) & 0.0690/-267.13 & 0.814 $\pm$ 0.096 & 2.07 $\pm$ 0.33 
& 0.218 $\pm$ 0.049 & 60.7 $\pm$ 6.9 \nl
\tablenotetext{a}{from (1) Bruzual \& Charlot 1993, (2) Rocca-Volmerange \& Guiderdoni
1988, (3) Poggianti 1997 $\Omega = 1.0$ and (4) $\Omega = 0.2$, (5) Pozzetti et al. 1996}
\tablenotetext{b}{Glazebrook et al. 1995}
\tablenotetext{c}{Mobasher et al. 1993}
\enddata
\end{deluxetable}

\clearpage

\begin{figure}[h]
\plotone{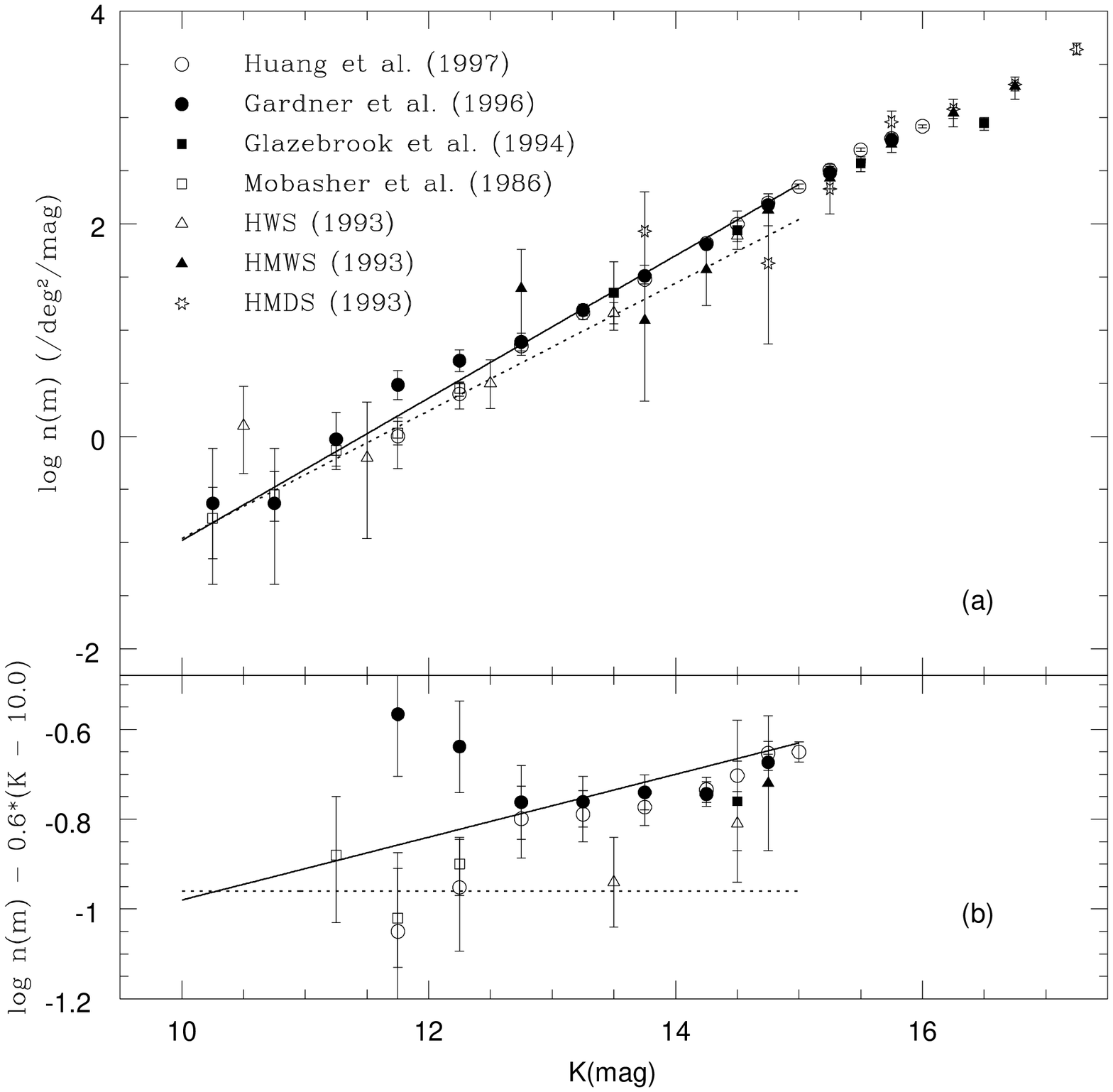}
\end{figure}

\clearpage

\begin{figure}[h]
\plotone{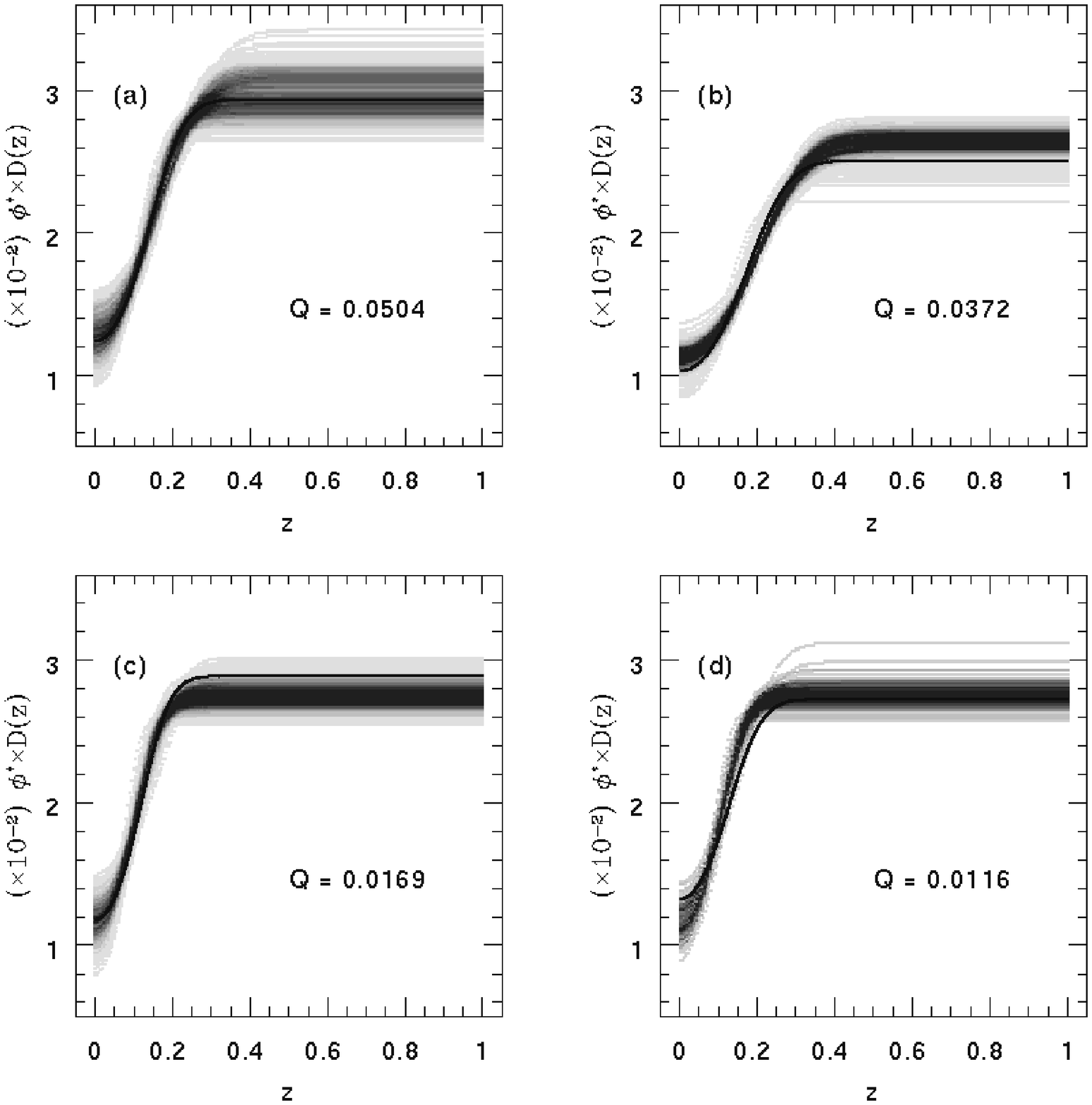}
\end{figure}

\clearpage

\begin{figure}[h]
\plotone{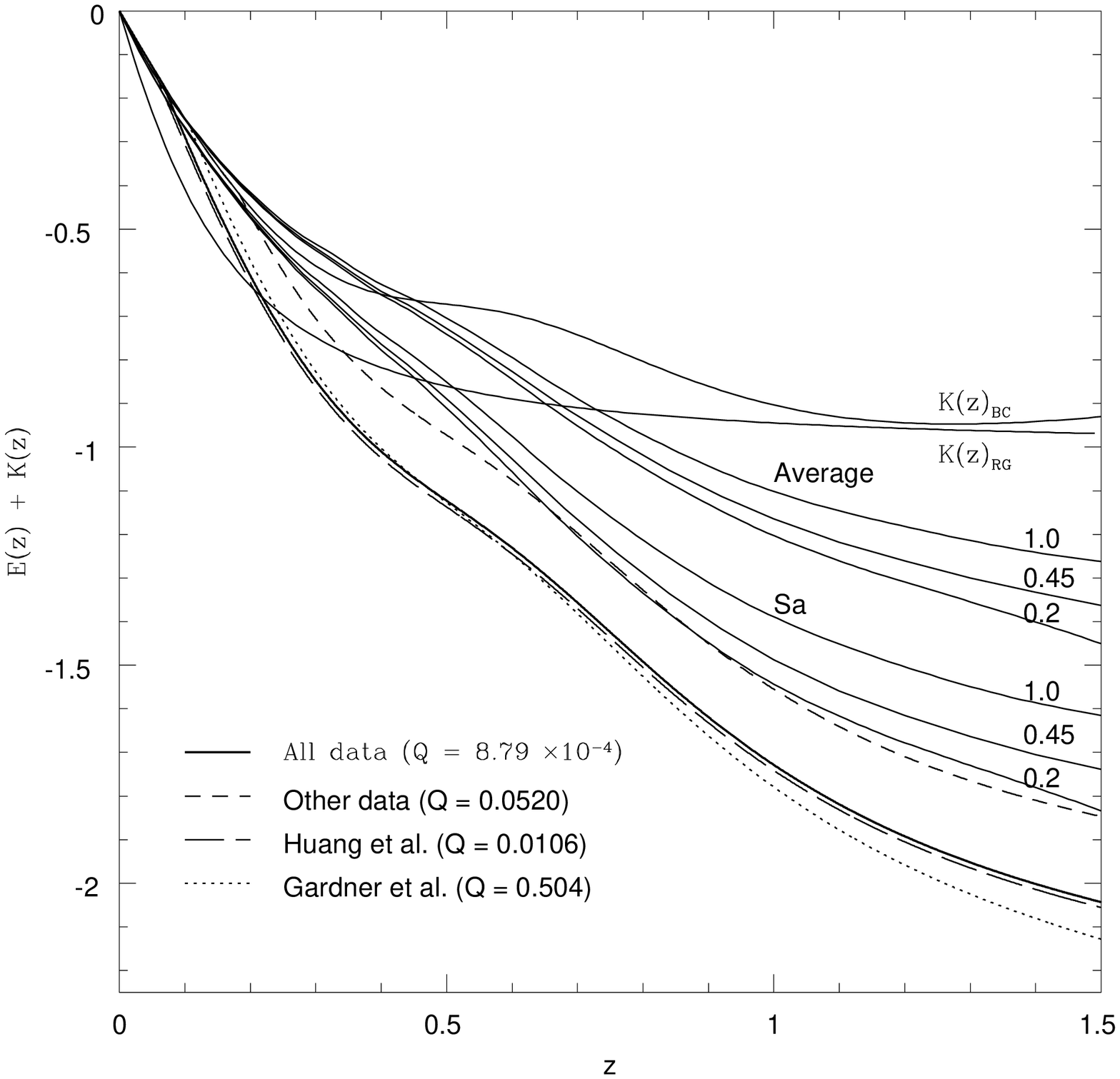}
\end{figure}

\end{document}